\documentstyle[fullpage,graphics]{article}

\newcommand{\hqp}{H^{qp}}
\newcommand{\vsc}{V_{sc}}
\newcommand{\vxc}{V_{xc}}
\newcommand{\sig}{\Sigma}
\newcommand{\Hop}{\hat{H}}
\newcommand{\Vop}{\hat{V}}
\newcommand{\hqpop}{\Hop^{qp}}
\newcommand{\vscop}{\Vop_{sc}}
\newcommand{\vxcop}{\Vop_{xc}}
\newcommand{\sigop}{\hat{\Sigma}}
\newcommand{\vc}{v_c}
\newcommand{\edft}{\varepsilon^{dft}}
\newcommand{\eqp}{\varepsilon}
\newcommand{\eps}{\epsilon}
\newcommand{\epsinv}{\eps^{-1}}
\newcommand{\ket}[1]{|#1\rangle}
\newcommand{\bra}[1]{\langle#1|}
\newcommand{\acv}[1]{A^S_{#1}}
\newcommand{\delr}{\partial_R}
\newcommand{\hbse}{H^{BSE}}

\newcommand{\prl}{Phys. Rev. Lett.}
\newcommand{\prb}{Phys. Rev. B}
\newcommand{\rmp}{Rev. Mod. Phys.}

\title{Excited-state Forces within a First-principles Green's Function
Formalism}

\author{Sohrab Ismail-Beigi and Steven G. Louie\\ Department of
Physics, University of California, Berkeley, CA 94720, and\\ Materials
Sciences Division, Lawrence Berkeley National Laboratory, Berkeley, CA
94720}

\date{\today}

\begin{document}

\maketitle

\begin{abstract}
We present a new first-principles formalism for calculating forces for
optically excited electronic states using the interacting Green's
function approach with the GW-Bethe Salpeter Equation method.  This
advance allows for efficient computation of gradients of the
excited-state Born-Oppenheimer energy, allowing for the study of
relaxation, molecular dynamics, and photoluminescence of excited
states.  The approach is tested on photoexcited carbon dioxide and
ammonia molecules, and the calculations accurately describe the
excitation energies and photoinduced structural deformations.
\end{abstract}

\vspace{0.5cm}

Calculation of optically excited electronic states and spectra has
recently become possible through use of the interacting two-particle
Green's function within the first-principles GW-Bethe Salpeter
Equation (GW-BSE) formalism~\cite{BSE,RL}, making possible the study
of photoinduced structural change and photoluminescence.  However,
within this methodology, one calculates optical properties at fixed
ionic positions: barring inefficient finite-difference schemes or
shrewd guesses, the direction in the multidimensional space of ionic
configurations best optimizing the geometry is unknown.  The problem
is of practical importance since structural changes due to optical
excitation are general phenomena which cause atomic rearrangement or
dissociation and change the structure or symmetry of defects.  The
possibility of efficient {\em ab initio} calculation of excited-state
forces opens new doors to reliable modeling and study of such changes.

To this end, we present a new formalism for calculating excited-state
forces within the GW-BSE approach.  We develop the theory for the
force calculations and the approximations requisite to render
computations tractable and then carry out practical tests for two
molecules.  We also compare our results to those of constrained
density functional theory (CDFT)~\cite{CDFT}.  Our development of
force calculations parallels recent quantum chemistry advances
(e.g. ~\cite{Stanton}) where analytical forces can be calculated for
excited-state methods (CIS, CIS(D), EOM-CCSD).  These methods provide
an overall accuracy similar to the GW-BSE
method~\cite{Stanton,Sattelmeyer}.  However, the GW-BSE method scales
as $N^4$ ($N$ is the number of atoms in the system) whereas these
methods scale as $N^6$ or worse.  In addition, quantum chemistry
methods are much more difficult to apply to the bulk properties of
solids whereas the GW-BSE method scales equally well in the bulk
limit.

Within the {\em ab initio} GW-BSE approach, the ground-state electron
density, total energy, forces, and single-particle states
$\ket{i}^{dft}$ and eigenvalues $\edft_i$ are obtained using density
functional theory (DFT).  We then construct the RPA dielectric
function $\epsinv$ and the screened interaction $W=\epsinv\vc$, where
$\vc$ is the Coulomb interaction $\vc(\vec{r},\vec{r}') =
1/|\vec{r}-\vec{r}'|$.  Quasiparticle excitations are found by using
the GW approximation to the self-energy, $\sig=iGW$~\cite{HL}.  The
effective quasiparticle Hamiltonian $\hqp$ is
\begin{equation}
\hqpop = \hat{T} + \vscop + (\sigop-\vxcop)\,,
\label{eq:hqp}
\end{equation}
where $T$ is the kinetic operator and $\vsc=V_{ion}+V_H+\vxc$ is the
sum of the ionic, Hartree, and DFT mean-field exchange-correlation
potentials.  We solve the Dyson equation $\hqpop\ket{i} = \eqp_i
\ket{i}$ to obtain quasiparticle energies $\eqp_i$ and eigenstates
$\ket{i}$.

Two-particle excited-state properties are obtained by solving the
Bethe-Salpeter equation of the two-particle Green's function.  We
employ the ``standard'' positive-frequency version of the BSE
eigenvalue equation and restrict to static screening~\cite{BSE,RL}.
The BSE eigenvalue equation is
\begin{eqnarray}
\sum_{c'v'} \hbse_{cv,c'v'} \acv{c'v'} & = & \Omega_S \, \acv{cv}\,,
\\ \mbox{with}\label{eq:bse} \ \ \ \ \ \ \ \ \ \ \hbse_{cv,c'v'} & = &
(\eqp_c-\eqp_v)\delta_{cc'}\delta_{vv'} + K_{cv,c'v'}\,.
\label{eq:hbse}
\end{eqnarray}
Here $c$ labels unoccupied (conduction) states and $v$ labels occupied
(valence) states, $\acv{cv}$ is the electron-hole amplitude for a
quasihole in state $v$ and a quasielectron in state $c$, $S$ labels an
excited state, $\Omega_S$ is the excitation energy, and $\hbse$ is the
effective electron-hole Hamiltonian.  The excited state energy is
given by $E_S = E_0 + \Omega_S$ where $E_0$ is the ground state
energy.  $K$ is the electron-hole interaction kernel with matrix
elements
\begin{equation}
K_{cv,c'v'} = \int c(1)^* v(2) \Xi(1234) c'(3) v'(4)^* \, d(1234) \,,
\label{eq:kdef}
\end{equation}
where $j(1)=\bra{1}j\rangle$.  The kernel $\Xi$ is
\begin{equation}
\Xi(1234) = -\delta(13)\delta(24)W(12) +
\delta(12)\delta(34)\vc(13)\,.
\label{eq:xidef}
\end{equation}
The static screening approximation means that $W=\epsinv\vc$ uses the
static dielectric function $\epsinv(\omega\!=\!0)$ in
Eq.~(\ref{eq:xidef}).  We note that the frequency dependent
$\epsinv(\omega)$ is used in the GW quasiparticle calculations.  We
use the standard normalization $\sum_{cv} |\acv{cv}|^2=1$.

Our aim is to compute excited-state forces, i.e. derivatives of $E_S$
versus the $3N$ ionic coordinates $R$, denoted as $\delr E_S$.  The
derivatives have two parts
\[
\delr E_S = \delr E_0 + \delr \Omega_S\,.
\]
DFT provides the ground-state derivatives $\delr E_0$~\cite{RMP}.  The
BSE provides expressions for $\Omega_S$, and we compute $\delr
\Omega_S$ directly.  Using Eq.~(\ref{eq:bse}) and the normalization
condition for $\acv{cv}$, we have
\begin{equation}
\delr \Omega_S = \sum_{cv,c'v'} {\acv{cv}}^* \acv{c'v'} \, \delr
\hbse_{cv,c'v'} \, ,
\label{eq:dromega}
\end{equation}
where
\begin{equation}
\delr \hbse_{cv,c'v'} = (\delr\eqp_c-
\delr\eqp_v)\delta_{cc'}\delta_{vv'} + \delr K_{cv,c'v'}\,.
\label{eq:drhbse}
\end{equation}
The derivative $\delr \hbse$ contains two types of terms, those
involving $\delr\eqp_i$ and those involving $\delr K$.  Below, we
adopt two physical approximations to render computations tractable.

Since $\hqpop\ket{i}=\eqp_i\ket{i}$ is a standard eigenvalue equation,
standard first order perturbation theory yields
\begin{eqnarray}
\delr\eqp_i & = & \bra{i}\delr\hqpop\ket{i}\,,\nonumber\\
P^R_{ji} & \equiv & \bra{j}\left\{\delr\ket{i}\right\} =
\left\{
\begin{array}{cc}
0 & \mbox{if } \eqp_i=\eqp_j\\
\frac{\bra{j}\delr\hqpop\ket{i}}{\eqp_i -\eqp_j} &
\mbox{if } \eqp_i\neq\eqp_j
\end{array}
\right\} .
\label{eq:pdef}
\end{eqnarray}
Evaluating $\delr\hqpop = \delr\vscop + \delr(\sigop-\vxcop)$ is
burdensome as $\delr \sigop$ contains the derivatives $\delr G$ and
$\delr\epsinv$, both prohibitive to calculate. We choose a physical
approximation: as discussed above, $\vsc$ contains the dominant ionic,
Hartree, and mean-field exchange-correlation potentials which bind
solids and molecules.  The weaker correction $\sig-\vxc$ consists
largely of a constant ``scissors-shift'' with a weak dependence on
$R$.  Therefore, we approximate $\delr\hqpop \cong \delr\vscop$.  To
calculate $\delr\vsc$, we employ density functional perturbation
theory~\cite{Gonze}: an auxiliary quadratic functional is minimized,
and at its minimum $\delr\vsc$ is easily found.  We perform $3N$
minimizations for the $3N$ choices of $R$.

Considering $\delr K$ and examining Eqs.~(\ref{eq:kdef}) and
(\ref{eq:xidef}), there are two distinct types of derivatives: those
containing $\delr \ket{i}$ and those containing $\delr W$.  For the
first set, we employ Eq.~(\ref{eq:pdef}) and sum over intermediate
states to reach convergence.  For the second set, we now argue (and
numerically verify below) that they are negligible. Specifically, in
the GW-BSE formalism, $W=\epsinv\vc$ where $\eps = I-\vc P$ and $P$ is
the polarizability. Within the RPA approximation that we employ,
$P=iGG$ is a function of $G$ alone and thus so are $\eps$ and $W$, so
the chain rule yields $\delr W(12) = \int\frac{\delta W(12)}{\delta
G(34)}\cdot\delr G(34) d(34)$. In deriving of the interaction kernel
$\Xi$, we assume that $\delta W / \delta G \cong 0$~\cite{RL}.
Judging from the success and accuracy of the BSE based on this
assumption, we conclude that we may set $\delr W\cong 0$.  Thus, we
arrive at the following expression for $\delr K$:
\begin{eqnarray}
\delr K_{cv,c'v'} & = & \sum_j \left[ {P^R_{jc}}^* \, K_{jv,c'v'} +
P^R_{jv} \, K_{cj,c'v'} + \right. \nonumber\\ & & \ \ \ \ \ \
\left. P^R_{jc'} \, K_{cv,jv'} + {P^R_{jv'}}^* \, K_{cv,c'j} \right]
\, .
\label{eq:delrK}
\end{eqnarray}
Taken together, Eqs.~(\ref{eq:dromega}-\ref{eq:delrK}) provide us with
an explicit expression for $\delr \Omega_S$ and hence $\delr E_S$.

We now present test applications to verify the accuracy of our
approach for molecules for which high quality excited-state
observations are available.  We begin by considering the first singlet
excited state of carbon monoxide (CO) which has one degree of freedom
$R$.

We carry out density functional calculations using the plane-wave
pseudopotential method within the local density approximation
(LDA)~\cite{RMP}. We use Kleinmann-Bylander pseudopotentials with $s$
and $p$ projectors ($p$ local) for C and O with cutoff radii of
$r_c=1.3$ a.u.~\cite{KB} and expand the electronic states to a plane
wave cutoff of 70 Ry.  Our periodic supercell is a $7\times 7\times 7$
\AA$^3$ cube.  We sample the Brillouin zone at $k=0$. We perform GW
calculations using the generalized plasmon-pole model~\cite{HL}. We
include 600 bands in the GW calculations to converge {\em absolute}
quasiparticle and ionization energies.  In the GW-BSE computations, we
truncate the Coulomb interaction beyond 3.5 \AA\ to avoid spurious
periodic image interactions~\cite{RL}.  With these parameters, all
reported energies are converged to 0.05 eV.  For comparison, we also
calculate excited-state properties with the constrained-LDA (CLDA)
method: we occupy the LUMO with an electron taken from the HOMO and
iterate to self-consistency.

We mention two technical issues: (i) off-diagonal matrix elements of
$\hqp$ in the DFT basis are required: their neglect changes $\Omega_S$
by $\sim\!\pm0.2$eV, as noted previously~\cite{QMCBSE}; (ii) in the
DFT supercell calculations, the Coulomb interaction has infinite
range: this creates an ambiguity in the vacuum level and shifts
quasiparticle energies by a constant.  By varying the supercell volume
$V$, we find the shift proportional to $1/V$ and extrapolate to
$V\rightarrow\infty$.

\begin{figure}
\resizebox{3.0in}{!}{\includegraphics{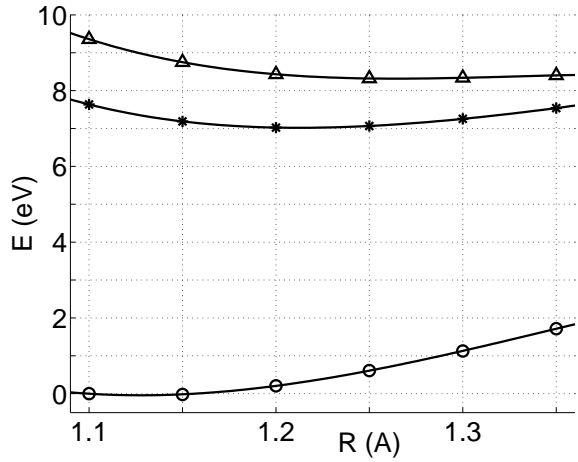}}
\caption{$X^1\Sigma^+$ ground-state LDA (circles) and $A^1\Pi$ first
excited-state energies (triangles are GW-BSE, stars are CLDA) versus
the bond length $R$ for CO.  The continuous curves are polynomial fits.}
\label{fig:CO}
\end{figure}

Fig.~\ref{fig:CO} displays the energy of the ground state and first
excited state of CO as a function of the bond length.
Table~\ref{table:CO} lists calculated and experimental properties of
the ground state ($X^1\Sigma^+$) and excited state ($A^1\Pi$):
equilibrium bond lengths $R_e$, harmonic vibration frequency
$\omega_e$, ionization energy IP (the HOMO energy for the LDA), and
the minimum-to-minimum transition energy $T_e$.  Results are also
presented for representative quantum chemical methods for which forces
have been developed.  The ground-state LDA $R_e$ and $\omega_e$ are in
good agreement with experimental results, and the quantum chemical
results are slightly superior due to the better treatment of
correlation. While the LDA HOMO energy lies far from the ionization
energy, the GW results remove this error.  The BSE results for the
transition energy $T_e$ to the excited state agree with experiment
with an error typical of the method~\cite{RL} whereas the CLDA
transition energy is off by 1 eV.  For $R_e$ and $\omega_e$ both the
CLDA and BSE perform equally well; while we do not have an explanation
for the size of the error, the fact that for this particular
excitation the CLDA produces an $\omega_e$ of similar quality to the
BSE one is explained by the observation (see below) that both methods
produce similar variation of force versus bond length.  The BSE
results for $T_e$ are of equal or better quality than the quantum
chemical results whereas for $R_e$ and $\omega_e$ they are slightly
worse.

\begin{figure}
\resizebox{3.0in}{!}{\includegraphics{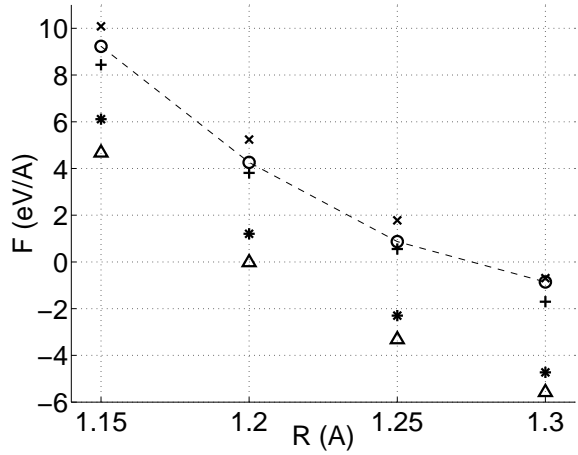}}
\caption{Absolute magnitudes of $A^1\Pi$ excited state forces for
CO. Circles are ``exact'' GW-BSE forces. Crosses and pluses are GW-BSE
forces on C and O. Triangles are GW-BSE forces with $\delr K=0$.
Stars are CLDA forces.  The dashed curve is a guide for the ``exact''
forces.}
\label{fig:fCO}
\end{figure}

Turning to the forces, we wish to know how accurately we can compute
$\delr E_S$, i.e. the slope of the curves in Fig.~\ref{fig:CO}.
Fig.~\ref{fig:fCO} presents the calculated forces, as formulated
above, along with ``exact'' forces obtained from two-point finite
differences of the BSE energies.  We also compute forces by assuming
$\delr K=0$ in Eq.~(\ref{eq:drhbse}), a quasiparticle-only treatment
and the closest analogue to the single-particle CLDA.  Both these
single-particle methods predict equal and opposite forces for the C
and O atoms, signaling that $\delr\hqp\cong\delr\vsc$ is a good
approximation.  Interestingly, both produce similar forces that depart
from the ``exact'' forces by essentially a constant.  Hence, both
methods predict well the variation of the force versus $R$, an {\em a
posteriori} verification of our physical intuition that changes in the
mean-field potential $\vsc$ dominate.

When we include contributions from $\delr K$, the forces improve
markedly.  Since we assume that $\delr W\cong 0$, the forces on C and
O are no longer exactly equal and opposite.  However, as
Fig.~\ref{fig:fCO} shows, the deviations are quite small on the
relevant scale: as remarked above, the approximation $\delr W\cong 0$
is excellent in practice.  If we subtract the unphysical net force on
the center of mass (i.e. averaging the C and O values in
Fig.~\ref{fig:fCO}), the calculated force become essentially
``exact''.

CO has a single degree of freedom allowing for careful study.
However, calculation of forces is truly useful when there are many
degrees of freedom and one does not know, {\em a priori}, which are
the relevant ones for a given excited state.  We now consider the
first singlet excited state of ammonia, NH$_3$, a molecule with six
degrees of freedom.  We employ identical methods as in the case of CO
and provide key parameters: $s$ projector for H ($r_c$=0.8 a.u.); $s$
and $p$ projectors for N ($p$ local, $r_c$=1.0 a.u.); identical
supercell, k-points, and Coulomb truncation radius; a 50 Ry cutoff for
the wave functions; 600 bands in the GW portion; all energies
converged to 0.05 eV.

Table~\ref{table:NH3} lists properties of the LDA ground state.
Starting with this ground state, for the BSE we consider the first
singlet state, and for the CLDA we promote an electron from HOMO to
LUMO.  We compute excited state forces and perform relaxations until
bond lengths are converged to 0.01 \AA, bond angles to 1$^o$, and
transition energies to 0.01 eV.  Table~\ref{table:NH3} presents
results for the relaxed excited state using CLDA, GW-BSE, and
representative quantum chemical methods.  The BSE excitation energy
compares well with experimental and quantum chemical values. (The
flattening of NH$_3$ is along the famous ``umbrella'' mode.)

Based on these two cases, we have verified the accuracy of the BSE
method for calculation of excited state energies and geometries.  Our
approach provides excited-state forces which are, to an excellent
approximation, the derivatives of the BSE energies.  Intriguingly, for
these two cases, CDFT yields inferior excitation energies but predicts
geometries of comparable quality to the BSE ones.  While encouraging,
there are a number of serious problems with wider use of CDFT.  Use of
CDFT is straightforward when the excited state is composed mainly of a
single configuration: in the cases above, the HOMO-LUMO combination
has probability above 90\% for the excited state.  However, for higher
excited states or larger systems, we can have multiple configurations,
something we can not know without solving the BSE.  Therefore, we
believe that CDFT may be a useful guide in certain circumstances, but
results thus obtained must be carefully tested by more sophisticated
methods.

In brief, we present an {\em ab initio} formalism for calculating
excited-state forces within the GW-BSE method as well as
approximations allowing for computational tractability.  We compute
the photoexcited properties of molecules and verify the accuracy of
(a) the GW-BSE formalism for describing the excited-state energies and
structural relaxations, and (b) the forces as per our formalism.  The
calculations are as accurate as leading methods used in quantum
chemistry (for which analytical force calculations are available)
while scaling significantly better with system size and being easily
applicable to the bulk.

This work was supported by NSF grant \#DMR-0087088 and by the Office
of Energy Research, Office of Basic Energy Sciences, Materials Science
Division of the U.S. DOE contract \#DE-AC03-76SF00098.  Computer
resources were provided by the DOE at the Lawrence Berkeley National
Laboratory National Energy Research Scientific Computing Center and by
the NSF National Partnership for Advanced Computational Infrastructure
at the San Diego Supercomputing Center.  This work was facilitated by
the DOE Computational Materials Science Network.

\begin{table}
\begin{tabular}{cccc}
& & Ground state ($X^1\Sigma^+$)\\
 & $R_e$ (\AA) & $\omega_e$ (cm$^{-1}$) & IP (eV)\\
LDA & 1.13 & 2050 & 9.1\\
GW & -- & -- & 14.1 \\ 
MP2~\cite{Stanton} & 1.133 & 2151 & -- \\
CCSD~\cite{Stanton} & 1.124 & 2243 & -- \\
Expt.~\cite{NIST} & 1.128 & 2170 & 14.01\\
\\
& & Excited state ($A^1\Pi$) \\
& $R_e$ (\AA) & $\omega_e$ (cm$^{-1}$) & $T_e$ (eV)\\
CLDA & 1.21 & 1720 & 7.02\\
GW-BSE & 1.26 & 1290 & 8.32\\
CIS~\cite{Stanton} & 1.21 & 1633 & 8.83\\
EOM-CCSD~\cite{Stanton} & 1.22 & 1593 & 7.91\\
Expt.~\cite{NIST} & 1.24 & 1518 & 8.07
\end{tabular}
\caption{Ground state and excited state data for CO: equilibrium bond
length $R_e$, harmonic vibrational frequency $\omega_e$, ionization
potential IP, and $X^1\Sigma^+\rightarrow A^1\Pi$ minimum-to-minimum
transition energy $T_e$.}
\label{table:CO}
\end{table}

\begin{table}
\begin{tabular}{cccc}
& & Ground state ($\tilde{X}^1A_1$) \\
& $R_e$ (\AA) & $\theta$ ($^o$) & IP (eV)\\
LDA & 1.03 & 105.0  & 6.2\\
GW & -- & -- & 10.7 \\ 
Expt.~\cite{NIST,Herzberg} & 1.01 & 106.7 & 10.1\\
\\
& & Excited state ($\tilde{A}^1A_2''$)\\
 & $R_e$ (\AA) & $\theta$ ($^o$) & $T_e$ (eV)\\
CLDA & 1.08 & 120 & 5.05 \\
GW-BSE & 1.08 & 120 & 5.52 \\
CASSCF~\cite{casscf} & 1.06 & 120 & 5.49\\
CEPA~\cite{casscf} & 1.06 & 120 & 5.63\\
Expt.~\cite{Herzberg} & 1.08 & 120 & 5.7
\end{tabular}
\caption{Ground and excited state data for NH$_3$: equilibrium
bond length $R_e$, H-N-H angle $\theta$, ionization potential IP,
and $\tilde{X}^1A_1\rightarrow \tilde{A}^1A_2''$ transition energy
$T_e$.  The experimental $T_e$ contains a zero-point contribution of
unknown but presumably small size.}
\label{table:NH3}
\end{table}

\end{document}